# Non-Hermitian CP-Symmetric Dirac Hamiltonians with Real Energy Eigenvalues

A. D. Alhaidari[†]

*Saudi Center for Theoretical Physics, Jeddah 21438, Saudi Arabia*

**Abstract**: We present a large class of non-Hermitian non-PT-symmetric two-component Dirac Hamiltoninas with real energy spectra. These Hamiltonians are invariant under the combined action of "charge" conjugation (two-component transpose) and space-parity. Examples are given from the two subclasses of these systems having localized and/or continuum states with real energies.

## I. INTRODUCTION

In an introductory course to quantum mechanics, hermiticity (precisely, self-adjointness) is the property that one usually employs to show that the energy eigenvalues of the Hamiltonian are real [1]. Nonetheless, in a seminal work by Bender and Boettcher in 1998 [2], it was shown that there exist non-hermitian Hamiltonians with real energy spectra. Since then, the physics and math literature on the subject grew very rapidly and many examples were given for such systems. For a recent review, one may consult [3] and references therein. Soon after [2], it was shown that Hamiltonian operators associated with this class of problems have a common property. They are invariant under the combined action of space inversion ($\mathcal{P}$) and time reversal ($\mathcal{T}$). That is, such Hamiltonians are $\mathcal{PT}$-symmetric. The work on this subject was not limited to the nonrelativistic theory. In fact, Non-Hermitian Dirac Hamiltonians with real relativistic energy spectra were also found [3-11]. All studies in this area were limited to the $\mathcal{PT}$-symmetric class of problems. In this work, however, we show that there exists a large class of non-hermitian non-$\mathcal{PT}$-symmetric two-component Dirac Hamiltoninas with real energy eigenvalues. The Hamiltonian matrix operators in this class are invariant under the combined action of "charge" conjugation and space inversion. We formulate the problem in the following section and then give analytic and numerical examples in Sec. III.

## II. FORMULATION

The Dirac equation in n+1 space-time for a particle, of rest mass $m$, under the influence of an external potential matrix $\mathcal{V}(t,x)$ is obtained by variation of the invariant action, $\int \mathcal{L}(x)d^{n+1}x$, whose Lagrangian reads as follows

---
[†] Present Address: 211 Keech Drive, Redwood City, CA 94065

$$\mathcal{L} = i\hbar c \bar{\psi}\not{\partial}\psi - mc^2 \bar{\psi}\psi + \bar{\psi}\mathcal{V}\psi, \qquad (1)$$

where $\not{\partial} = \sum_\mu \gamma^\mu \partial_\mu$ is the Dirac operator, $\{\gamma^\mu\}_{\mu=0}^n$ are constant square matrices (the Dirac gamma matrices), $\partial_\mu = \left(\frac{1}{c}\frac{\partial}{\partial t}, \vec{\nabla}\right)$ is the (n+1)-gradient, and $\bar{\psi} = \psi^\dagger \gamma^0$. The Dirac matrices satisfy the anti-commutation relation $\{\gamma^\mu, \gamma^\nu\} = 2\mathcal{G}^{\mu\nu}$, where the n+1 Mikowski space-time metric is $\mathcal{G} = \text{diag}(+--\cdots)$. The resulting Dirac equation (in the conventional relativistic units $\hbar = c = 1$) is

$$\left(i\gamma^\mu \partial_\mu - m + \mathcal{V}\right)\psi = 0. \qquad (2)$$

In 1+1 space-time, we choose the representation of the Dirac gamma matrices as $\gamma^0 = \sigma_3 = \begin{pmatrix} 1 & 0 \\ 0 & -1 \end{pmatrix}$ and $\gamma^1 = i\sigma_1 = i\begin{pmatrix} 0 & 1 \\ 1 & 0 \end{pmatrix}$, which makes the potential $\mathcal{V}(t,x)$ a 2×2 matrix and gives the following most general Dirac equation in 1+1 dimension

$$i\partial_t \begin{pmatrix} \psi_+ \\ \psi_- \end{pmatrix} = \begin{pmatrix} m+S+V & \partial_x + W \\ -\partial_x + U & -m-S+V \end{pmatrix} \begin{pmatrix} \psi_+ \\ \psi_- \end{pmatrix} = \begin{pmatrix} & H_D & \end{pmatrix} \begin{pmatrix} \psi_+ \\ \psi_- \end{pmatrix}, \qquad (3)$$

where $\{S, V, W, U\}$ are four independent space-time functions. The Dirac Hamiltonian matrix is Hermitian (i.e., $H_D^\dagger = H_D$) if and only if $W = U^* = P + iQ$ and all potential functions $\{S, V, P, Q\}$ are real. In that case, the interaction Lagrangian becomes

$$\bar{\psi}\mathcal{V}\psi = S(\bar{\psi}\psi) + A_\mu(\bar{\psi}\gamma^\mu\psi) - P(\bar{\psi}\gamma^5\psi), \qquad (4)$$

where $\gamma^5 = i\gamma^0\gamma^1 = \begin{pmatrix} 0 & -1 \\ 1 & 0 \end{pmatrix}$ and the potential components carry well-defined irreducible representations of the Lorentz group. The scalar, pseudo-scalar, and vector potentials are S, P, and $A_\mu = (V, Q)$, respectively. Moreover, the space-component of the vector potential, Q, could also be eliminated by a U(1) gauge transformation. However, from now on, we take $\{S, V, W, U\}$ in Eq. (3) to be real space-time functions. Thus, the Dirac Hamiltonian is self-adjoint (precisely, Hermitian) if and only if $U = W$. This is because the two parts of the Lagrangian relevant to this issue are

$$-W(\bar{\psi}\gamma_+\psi) - U(\bar{\psi}\gamma_-\psi) = W(\psi^\dagger \sigma_+ \psi) + U(\psi^\dagger \sigma_- \psi), \qquad (5)$$

where $\gamma_\pm = \frac{1}{2}(\gamma^5 \pm i\gamma^1)$, $\sigma_+ = \begin{pmatrix} 0 & 1 \\ 0 & 0 \end{pmatrix}$ and $\sigma_- = \begin{pmatrix} 0 & 0 \\ 1 & 0 \end{pmatrix}$. On the other hand, let us examine the following transformed Dirac Hamiltonian matrix

$$\gamma^5 H_D \gamma^5 = \begin{pmatrix} m+S+V & -\partial_x + U \\ \partial_x + W & -m-S+V \end{pmatrix}. \qquad (6)$$

Thus, $\gamma^5$ transposes the Hamiltonian matrix (without complex conjugation). It also acts to exchange the two components of the wavefunction as: $\psi_\pm \to \mp\psi_\mp$. In fact, $\gamma^5$ has all the properties of the charge conjugation matrix, $\mathcal{C}$. That is, $\mathcal{C}^T = \mathcal{C}^\dagger = \mathcal{C}^{-1} = -\mathcal{C}$ and $\mathcal{C}\gamma^\mu \mathcal{C}^{-1} = \pm(\gamma^\mu)^T$ [12]. However, one may not be at liberty to extend this technical analogy too far into the physics to assume conjugation of electric charges. Nonetheless, from now on we replace the designation $\gamma^5$ by $\mathcal{C}$. Applying the space-parity operator ($\mathcal{P}: x \to -x$) on (6) gives

$$\mathcal{CP} H_D \mathcal{CP} = \begin{pmatrix} m + \hat{S} + \hat{V} & \partial_x + \hat{U} \\ -\partial_x + \hat{W} & -m - \hat{S} + \hat{V} \end{pmatrix}, \tag{7}$$

where the caret over the function means $\hat{f}(t,x) = f(t,-x)$ and we have used the fact that space-parity and charge conjugation commute ($\mathcal{CP} = \mathcal{PC}$). Therefore, we conclude that $\mathcal{CP} H_D \mathcal{CP} = H_D$ if and only if: $S(t,-x) = S(t,x)$, $V(t,-x) = V(t,x)$, and $U(t,-x) = W(t,x)$. For example, if all the potential functions are even in $x$, then $U = W$ and $H_D$ is trivially Hermitian with $H_D = \mathcal{CP} H_D \mathcal{CP} = H_D^\dagger$. However, if $W$ is odd, then $U = -W$ and the Dirac Hamiltonian is not Hermitian but still satisfies the condition $\mathcal{CP} H_D \mathcal{CP} = H_D$. To be specific, we consider in this work the class of problems where $S = V = 0$ and $W(t,-x) \neq W(t,x)$ giving the non-Hermitian Dirac Hamiltonian

$$H_D = \begin{pmatrix} m & \partial_x + W(t,x) \\ -\partial_x + W(t,-x) & -m \end{pmatrix}, \tag{8}$$

which is $\mathcal{CP}$-symmetric. Then, the Dirac equation (3) for this class of problems reads as follows in components form

$$i\partial_t \psi_+ = m\psi_+ + \partial_x \psi_- + W(t,x)\psi_-, \tag{9a}$$
$$i\partial_t \psi_- = -m\psi_- - \partial_x \psi_+ + W(t,-x)\psi_+. \tag{9b}$$

It is worth noting that the Dirac Hamiltonian (8) is not $\mathcal{PT}$-symmetric. That is, $\mathcal{PT} H_D \mathcal{PT} \neq H_D$, where $\mathcal{T}$ is the time-reversal operator, which has the effect of complex conjugation, $i \to -i$. It is worth mentioning that the $\mathcal{CPT}$-symmetry being discussed in the quantum mechanics literature regarding non-Hermitian Hamiltonians with real spectra (including the present work) should not be identified with the fundamental $\mathcal{CPT}$-symmetry discussed in quantum field theory and particle physics [12].

Now, for time independent potential functions, $W(x)$, we can write $\psi_\pm(t,x) = e^{-i\varepsilon t}\chi_\pm(x)$, where $\varepsilon$ is the system's energy. In that case, Eq. (9) gives the following relations between the two components of the wavefunction

$$\chi_+(x) = \frac{1}{\varepsilon - m}\left[W(x) + \frac{d}{dx}\right]\chi_-(x), \quad \varepsilon \neq m \text{ (negative energy)}, \tag{10a}$$

$$\chi_-(x) = \frac{1}{\varepsilon + m}\left[W(-x) - \frac{d}{dx}\right]\chi_+(x), \quad \varepsilon \neq -m \text{ (positive energy)}, \tag{10b}$$

resulting the following second order differential equations

$$\left[\frac{d^2}{dx^2} + (W - U)\frac{d}{dx} - WU - U' + \varepsilon^2 - m^2\right]\chi_+ = 0, \tag{11a}$$

$$\left[\frac{d^2}{dx^2} + (W - U)\frac{d}{dx} - WU + W' + \varepsilon^2 - m^2\right]\chi_- = 0, \tag{11b}$$

where $U(x) = W(-x)$ and the prime stands for the derivative with respect to $x$. The question now is as follows: Are there real potential functions, $W(x)$, that result in non-zero physical solutions (bound and/or continuous) of Eq. (11) for real energies $\varepsilon$? We will show (by example) in the following section that the answer is affirmative.

For positive energies, one solves Eq. (11a) to obtain $\chi_+$ then substitute that in Eq. (10b) to obtain $\chi_-$. On the other hand, for negative energies, one solves Eq. (11b) to obtain $\chi_-$ then substitute that in Eq. (10a) to obtain $\chi_+$. It should be emphasized that the solution of Eq. (11a) does NOT belong to the same energy subspace as that of Eq. (11b) and the complete solution space is a union of these two subspaces. In the following section, we illustrate our findings by giving various examples (analytic and numerical) of bound and continuum solutions of Eq. (11) for real energies.

## III. EXAMPLES

If we define $\rho(x) = \int_x [W(y) - W(-y)] dy$ and write $\chi_\pm(x) = e^{-\frac{1}{2}\rho(x)} \phi_\pm(x)$, then Eq. (11) becomes

$$\left[ \frac{d^2}{dx^2} - R^2 \mp R' + \varepsilon^2 - m^2 \right] \phi_\pm(x) = 0, \qquad (12)$$

where $R(x) = \frac{1}{2}[W(x) + W(-x)]$. For bound state, $\lim_{|x| \to \infty} e^{-\frac{1}{2}\rho(x)} = 0$, whereas for scattering states $\lim_{|x| \to \infty} e^{-\frac{1}{2}\rho(x)}$ is finite. The effective potential associated with $\phi_\pm(x)$ is $V_\pm(x) = \frac{1}{2}(R^2 \pm R')$, which has the structure of the superpotentials in supersymmetric quantum mechanics (SSQM) [13]. Thus, besides performing the integration to find $\rho(x)$, all the tools of SSQM and shape invariance are at our disposal to find the solution of Eq. (12). We divide the class of solutions of the Dirac equation (9) into two. The odd parity subclass associated with the potential functions $W(-x) = -W(x)$ and a larger subclass without definite parity, which is associated with $W(-x) \neq \pm W(x)$.

### A. Examples in the subclass $W(-x) = -W(x)$

In these cases $V_\pm(x) = 0$ and $\chi_\pm(x) = e^{-\frac{1}{2}\rho(x)} \left( A_\pm e^{ikx} + B_\pm e^{-ikx} \right)$, where $k = \sqrt{\varepsilon^2 - m^2}$. For $|\varepsilon| < m$, the oscillatory factor becomes growing and decaying exponentials of which we choose the latter. As a particular example, we take $W(x) = A(\mu x)^{2n+1}$ where $A$ and $\mu$ are real parameters such that $\mu A > 0$, and $n$ is a non-negative integer. In such a case, we obtain bound states with $\rho(x) = \frac{A/\mu}{n+1}(\mu x)^{2(n+1)}$. Another example is when $W(x) = A \sinh(\mu x)$ with $\mu A > 0$, then we obtain the bound states $\chi_\pm(x) = e^{-\frac{A}{\mu}\cosh(\mu x)} \left( A_\pm e^{ikx} + B_\pm e^{-ikx} \right)$. On the other hand if $W(x) = A \sin(\mu x)$, then we obtain the continuum states $\chi_\pm(x) = e^{\frac{A}{\mu}\cos(\mu x)} \left( A_\pm e^{ikx} + B_\pm e^{-ikx} \right)$. Table 1 is a sample list of problems in this subclass.

## B. Examples in the subclass $W(-x) \neq \pm W(x)$

If we take $W(x) = A e^{-\mu x}$ such that $\mu A < 0$, then we obtain bounded solutions for all real energies with $\chi_\pm(x) = e^{\frac{A}{\mu}\cosh(\mu x)} \phi_\pm(x)$ and Eq. (12) becomes

$$\left[\frac{d^2}{dx^2} - A^2 \sinh^2(\mu x) \mp \mu A \sinh(\mu x) + \varepsilon^2 - M^2\right]\phi_\pm(x) = 0, \qquad (13)$$

where $M^2 = m^2 + A^2$. This is a Schrödinger equation for $\phi_\pm(x)$ with the confining potential $V_\pm(x) = \frac{1}{2}\left[A\sinh(\mu x) \pm \frac{\mu}{2}\right]^2 - \frac{\mu^2}{8}$. Thus, we expect that all real energy solutions be bounded. A second example is for $W(x) = A(\mu x)e^{-\mu x}$ with $\mu A > 0$ giving $\rho(x) = 2\frac{A}{\mu}[(\mu x)\sinh(\mu x) - \cosh(\mu x)]$ and $\phi_\pm(x)$ is a solution of Schrödinger equation with the effective potential $V_\pm(x) = \frac{1}{2}(R^2 \pm R')$, where $R(x) = A(\mu x) \times \cosh(\mu x)$. In Fig. 1 and Fig. 2, we show few plots of $\chi_\pm(x)$ from this subclass, which are obtained numerically for the given set of physical parameters. Table 2 is a sample list of problems in this subclass.

## ACKNOWLEDGEMENTS

The Author is grateful for the support granted by the Saudi Center for Theoretical Physics (SCTP). We also appreciate the help in literature search provided by S. Al-Marzoug and H. Bahlouli.

**TABLE CAPTION:**

**Table 1**: Sample list of $\mathcal{CP}$-symmetric problems in the subclass $W(-x) = -W(x)$ with $\chi_{\pm}(x) = e^{-\frac{1}{2}\rho(x)}\left(A_{\pm}e^{ikx} + B_{\pm}e^{-ikx}\right)$ and $k = \sqrt{\varepsilon^2 - m^2}$.

**Table 2**: Sample list of $\mathcal{CP}$-symmetric problems in the subclass $W(-x) \neq \pm W(x)$ with $\chi_{\pm}(x) = e^{-\frac{1}{2}\rho(x)}\phi_{\pm}(x)$ and $\phi_{\pm}(x)$ are solutions of $\left(\frac{d^2}{dx^2} - R^2 \mp R' + \varepsilon^2 - m^2\right)\phi_{\pm} = 0$.

**FIGURE CAPTION:**

**FIG. 1**: Plots of $\chi_{\pm}(x)$ for the case $W(x) = A e^{-\mu x}$. The upper (lower) component is shown with red (blue) color. The physical parameters are $m = 1$, $A = -1.0m$, and $\mu = 0.2m$. The boundary conditions are chosen as $\chi_{+}(0) = 1$ and $\chi_{-}(0) = 0$. The plots are given for (a) positive energy $\varepsilon = +2.0m$, and (b) negative energy $\varepsilon = -2.0m$.

**FIG. 2**: Plots of $|\chi(x)| = \sqrt{|\chi_{+}|^2 + |\chi_{-}|^2}$ for the case $W(x) = A\dfrac{1+(\mu x)}{1+(\mu x)^2}$. The physical parameters are $m = 1$, $A = 1.0m$, and $\mu = 0.5m$. The boundary conditions are chosen as $\chi_{+}(0) = 1$ and $\chi_{-}(0) = 0$. The plots are given for (a) positive energy $\varepsilon = +1.5m$, and (b) negative energy $\varepsilon = -1.5m$.

**Table 1**

| $W(x)$ | $\rho(x)$ | Remarks |
|---|---|---|
| $A(\mu x)^{2n+1}$ | $\frac{A/\mu}{n+1}(\mu x)^{2(n+1)}$ | $\mu A > 0$, localized |
| $A\sinh(\mu x)$ | $2\frac{A}{\mu}\cosh(\mu x)$ | $\mu A > 0$, localized |
| $A\sin(\mu x)$ | $-2\frac{A}{\mu}\cos(\mu x)$ | Non-localized |
| $A\sinh^{-1}(\mu x)$ | $2\frac{A}{\mu}\left[(\mu x)\sinh^{-1}(\mu x) - \sqrt{(\mu x)^2 + 1}\right]$ | $\mu A > 0$, localized |
| $A(\mu x)\ln(\mu x)^2$ | $\frac{A}{\mu}(\mu x)^2\left[\ln(\mu x)^2 - 1\right]$ | $\mu A > 0$, localized |

**Table 2**

| $W(x)$ | $\rho(x)$ | $R(x)$ | Remarks |
|---|---|---|---|
| $Ae^{-\mu x}$ | $-2\frac{A}{\mu}\cosh(\mu x)$ | $A\cosh(\mu x)$ | $\mu A < 0$, localized |
| $A(\mu x)e^{-\mu x}$ | $2\frac{A}{\mu}[(\mu x)\sinh(\mu x) - \cosh(\mu x)]$ | $-A(\mu x)\sinh(\mu x)$ | $\mu A > 0$, localized |
| $A(\mu x)^2 e^{-\mu x}$ | $-2\frac{A}{\mu}\left[(y^2+2)\cosh(y) - 2y\sinh(y)\right]$ | $Ay^2\cosh(y)$ | $\mu A < 0$, $y = \mu x$, localized |
| $A\dfrac{1+(\mu x)}{1+(\mu x)^2}$ | $\frac{A}{\mu}\ln\left[1+(\mu x)^2\right]$ | $\dfrac{A}{1+(\mu x)^2}$ | $\mu A > 0$, localized |

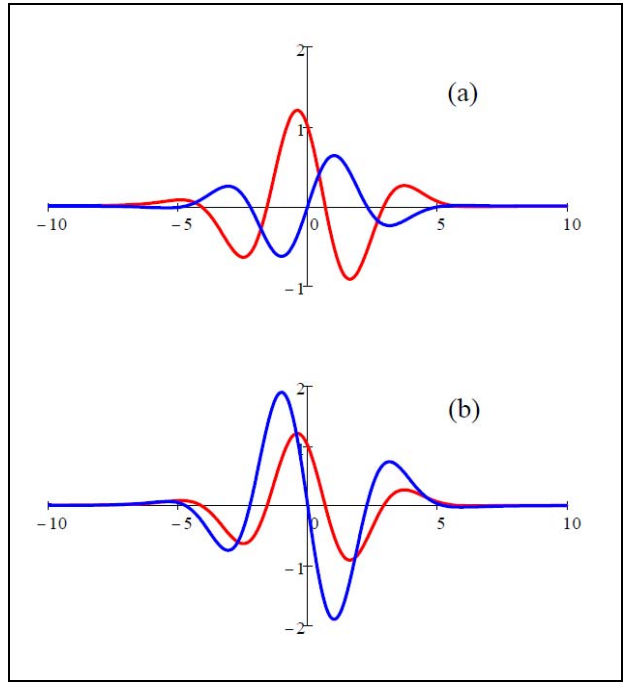

Fig. 1

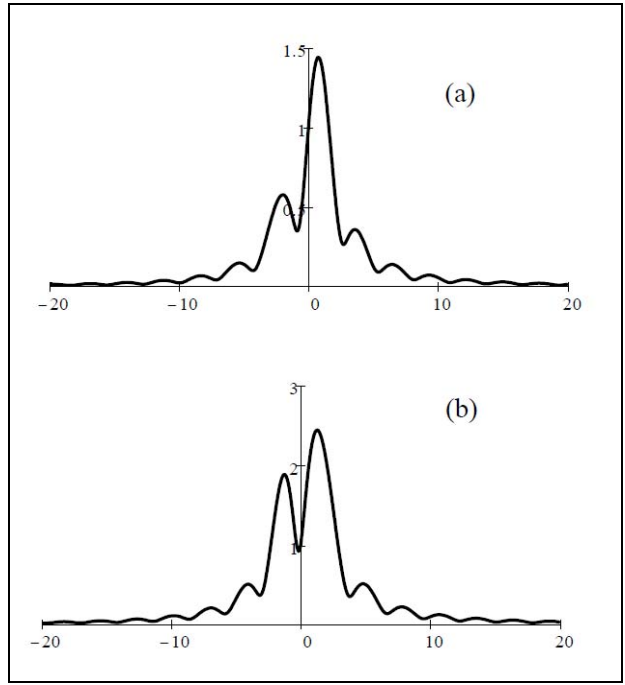

Fig. 2